\documentclass{ws-ijbc}
\usepackage{graphicx}
\usepackage{epstopdf}
\usepackage{float}
\usepackage{epsfig}
\usepackage{natbib}
\usepackage{amsmath}
\usepackage{times}
\usepackage{subfigure}
\usepackage{graphicx}
\usepackage{color}
\usepackage{dsfont}
\usepackage[utf8]{inputenc} 
\usepackage{verbatim}
\usepackage{xcolor}
\begin{document}

\catchline{}{}{}{}{} 

\markboth{D. Valle, A. Wagemakers, A. Daza and M.A.F Sanju\'{an}}{Characterization of Fractal Basins Using Deep Convolutional Neural Networks}

\title{CHARACTERIZATION OF FRACTAL BASINS USING DEEP CONVOLUTIONAL NEURAL NETWORKS}

\author{DAVID VALLE}
\address{Nonlinear Dynamics, Chaos and Complex Systems Group, Departamento de F\'{i}sica, Universidad Rey Juan Carlos, Tulip\'{a}n s/n, 28933 M\'{o}stoles, Madrid, Spain.\\
david.valle@urjc.es}

\author{ALEXANDRE WAGEMAKERS}
\address{Nonlinear Dynamics, Chaos and Complex Systems Group, Departamento de F\'{i}sica, Universidad Rey Juan Carlos, Tulip\'{a}n s/n, 28933 M\'{o}stoles, Madrid, Spain.\\
alexandre.wagemakers@urjc.es}

\author{ALVAR DAZA}
\address{Nonlinear Dynamics, Chaos and Complex Systems Group, Departamento de F\'{i}sica, Universidad Rey Juan Carlos, Tulip\'{a}n s/n, 28933 M\'{o}stoles, Madrid, Spain.\\
alvar.daza@urjc.es}

\author{MIGUEL A.F. SANJU\'{A}N}
\address{Nonlinear Dynamics, Chaos and Complex Systems Group, Departamento de F\'{i}sica, Universidad Rey Juan Carlos, Tulip\'{a}n s/n, 28933 M\'{o}stoles, Madrid, Spain.\\
Department of Applied Informatics, Kaunas University of Technology, Studentu 50-415, Kaunas LT-51368, Lithuania\\
miguel.sanjuan@urjc.es\footnote{Address for correspondence.}}

\maketitle

\begin{history}
\received{(to be inserted by publisher)}
\end{history}

\begin{abstract}
Neural network models have recently demonstrated impressive prediction performance in complex systems where chaos and unpredictability appear. In spite of the research efforts carried out on predicting future trajectories or improving their accuracy compared to numerical methods, not sufficient work has been done by using deep learning techniques in which they characterize the unpredictability of chaotic systems or give a general view of the global unpredictability of a system. In this work we propose a novel approach based on deep learning techniques to measure the fractal dimension of the basins of attraction of the Duffing oscillator for a variety of parameters. As a consequence, we provide an algorithm capable of predicting fractal dimension measures as accurately as the conventional algorithm, but with a computation speed about ten times faster. 
\end{abstract}


\section{Introduction} \label{sec:Introduction}

Despite being governed by deterministic rules, chaotic systems are hard to predict since the slightest disturbance can cause a completely different evolution of the system. This inherent property of chaos is closely related to fractal geometry, so it is not surprising that both disciplines were born almost simultaneously~\cite{peitgen1992chaos}. One example where the relationship between chaos and fractal geometry becomes obvious is the basins of attraction of multistable dynamical systems~\cite{nusse1996basins, aguirre2009fractal}. In chaotic dynamical systems with multiple possible outcomes, the basin boundaries display complicated fractal structures at every scale. This hinders the long-term prediction of initial conditions starting in the vicinity of such fractal boundaries.\\

Basins can be represented as images where each pixel corresponds to an initial condition and the color represents the final state. One can clearly distinguish certain patterns in the basins, such as filaments or boundaries meandering over and over again. Therefore, due to the presence of these patterns, basins are objects well adapted to be analyzed by algorithms based on machine learning.\\

The use of machine learning and deep learning techniques opens a wide field of applications for the classification of basins of attraction. Indeed, machine learning has already been used to help solving problems in nonlinear dynamics such as improving the prediction time of chaotic trajectories ~\cite{fan2020long,pathak2018hybrid}, reconstructing the long-term dynamics of a system and therefore either its chaotic attractor or its basin of attraction ~\cite{lu2018attractor, basinapproximation}, as well as classifying data with associated fractal properties~\cite{kirichenko2018machine, shi2018signal}. All these, among many other applications such as the ones shown in ~\cite{tang2020introduction} constitute applications of machine learning algorithms into nonlinear dynamics and chaos.\\

However there is a lack of research on characterizing the general unpredictability of systems using machine learning algorithms. These algorithms are mainly used on the prediction of systems for a fixed set of parameters and therefore they do not provide a general overview of the global unpredictability of a system. Nevertheless, basins of attraction give information of the unpredictability of a system that can be quantified by analyzing the fractal dimension associated to the basin boundaries. This quantity is commonly measured using the box-counting algorithm proposed by McDonald et al.~\cite{mcdonald1985fractal}. Our main goal is to reproduce this algorithm using machine learning techniques and therefore yields a tool based on deep learning able to accurately quantify the fractal dimension of any basin of attraction for a given system on a short period time.\\

There might be different ways to measure the fractal dimension of the basins with machine learning. A first option could be the use of Artificial Neural Networks (ANNs), which are able to detect patterns in images~\cite{bishop2006pattern}. ANNs are inspired by the nervous system and they are mainly formed of a high number of interconnected computational nodes, referred to as neurons. These neurons work in a distributed fashion to collectively learn  from the input using tunable parameters (weights and biases) in order to optimize its final output.\\

However, ANNs have serious limitations. First, the number of weights per neuron in the first layer is proportional to the number of pixels of the image. Furthermore, we would need to take into account the number of hidden layers in the architecture of the neural network and the number of epochs required to train the network. An excessive number of hidden layers or epochs would decrease the accuracy of the model due to overfitting data~\cite{bishop2006pattern}.\\

The previous problem can be circumvented by using Convolutional Neural Networks (CNNs). They are considered as deep learning tools based on learning data models. They were developed by LeCun et al.~\cite{726791} in 1998 as a class of deep feed‐forward ANN. They take nearby pixels from the input image and process them through multiple layers of convolutional filters based on a small kernel matrix. This convolutional kernel has a size much smaller than the presented image and it manages to extract features (such as edges or vertices) according to some weights and biases. These parameters are tuned during training depending whether the features that the kernel is extracting are improving or decreasing the accuracy of the neural network. Not only that, CNNs have demonstrated excellent performance at tasks such as hand-written digit classification and face detection. In the past years, several papers have shown that they can also deliver outstanding performance on more challenging visual classification and regression tasks ~\cite{ NIPS2012_c399862d, espectroscopia}. Several factors are responsible for this renewed interest in CNNs models such as: the availability of much larger training sets with millions of labeled examples; powerful GPU implementations, making the training of very large models practical; and better model regularization strategies such as Dropout ~\cite{understandingvision}. For this reason, we believe that convolutional neural networks are optimal to work on our regression problem. \\

As we have already mentioned, we specifically aim in this paper to quantify the unpredictability of a chaotic system for a variety of parameters using deep learning techniques. To do such task we measure the fractal dimension of the basin boundaries in a basin of attraction using a CNN trained via  supervised learning, that is, the CNN learns through pre-labeled inputs which act as targets. Using this setup we obtain a trained CNN capable of accurately predicting the fractal dimension of basins of attraction in the Duffing Oscillator, although not as accurately as the conventional method due to the fact that the error in this conventional method can be arbitrarily small, but with a computation time ten times faster than the conventional estimation method. Nevertheless, we prove how our algorithm is capable of estimating the fractal dimension of a basin of attraction, so that we could improve the accuracy of our convolutional neural network with a bigger dataset or subtle hyperparameter tuning until it reaches the accuracy and robustness of our ground truth but with the advantage of the fast computation time.\\

The rest of the paper is organized as follows. Section~\ref{ArchitectureChapter} shows the architecture, data augmentation techniques, and hyperparameters set for training the CNN. Section~\ref{FractalDimension} is devoted to explain the image set used, how we made the distribution between training, validation, and testing sets, and finally how well did our neural network during training. In section~\ref{Results}, we show the results of testing our trained CNN on the proposed testing set of images, and we compare our results with the conventional method proving how our method is more efficient. Finally, in section~\ref{Conclusions} we discuss our results and propose a way to improve the prediction performance of the CNN.

\section{Architecture of the CNN}
\label{ArchitectureChapter}

\begin{figure*}
\includegraphics[scale=0.20]{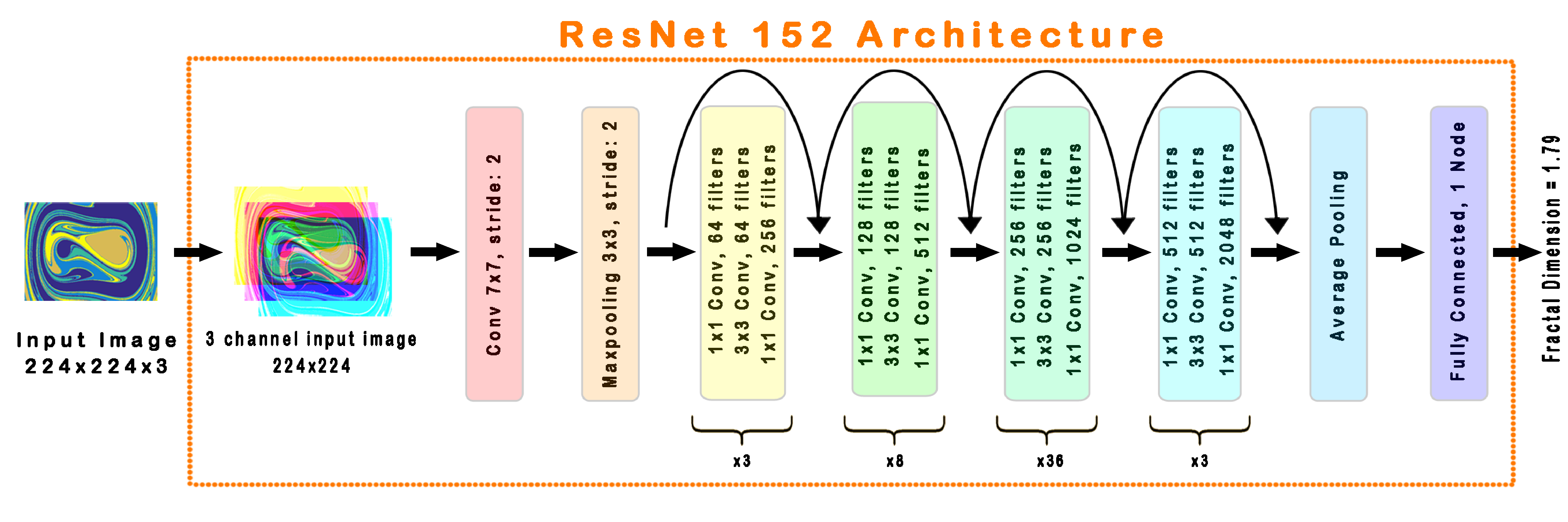}
\label{architecture}
\caption{Outline of ResNet152 used in this work. The input of the network is an $ 224 \times 224 \times 3$ image that gets splitted into its RGB channels. Then, a convolutional kernel of $7$ $\times$ $7$ convolutes over the images producing a matrix with a reduced number of parameters. As a result, a batch normalization layer and a ReLU activation function takes place. This batch normalization and activation occurs after every convolutional layer in the network and are not shown in the schematic representation. Then, the matrix enters the Maxpooling layer, where the network extracts the features that it considers important for the optimization of the model and therefore produces a new matrix. This new matrix goes into the main blocks of the ResNet152 where several convolutions (and hence normalizations and activations) take place. The numbers shown below the main blocks in the figure represent the times the block is repeated in the network. Once the convolutions are finished another pooling is done, and therefore the network extracts the values that are considered important for making predictions. Finally, the output matrix is flattened to a vector, and each element is connected through a fully connected layer to one node yielding a number representing the fractal dimension of the basin of attraction shown in the input image. It is worthy to mention that there are residual connections between each block in the architecture, minimizing the error as proven by Kaiming et al.~\cite{he2016deep}.}
\label{Architecture}
\end{figure*}

For convolutional neural networks, the prediction error increases after a certain number of layers are added. Kaiming et al.~\cite{he2016deep} solved this problem by using residual connections in the network, allowing deeper networks with a lower error percentage than those produced by classical architectures such as Vgg~\cite{simonyan2014very}, GoogLeNet~\cite{szegedy2015going}, or PReLU-net~\cite{he2015delving}. Here, we use the architecture of such neural network with residual connections (ResNet). The schematic representation of this architecture is shown in Fig.~\ref{Architecture} where we can see the entire process that the neural network goes through. First, the image is input into the neural network where is splitted into its RGB color channels, then each channel goes through the convolutional process where the neural network is able to extract features useful for the regression problem, and finally the last fully connected layer of the network outputs a number that is to be interpreted as the fractal dimension of the basin of attraction shown in the input image.\\

For this work, a ResNet152 architecture has been implemented with the Python programming language libraries: Tensorflow (2.5.0), and Keras (2.5.0). Also, since neural network training mainly requires GPUs, we have used an Nvidia Quadro RTX 5000. The Keras and Tensorflow libraries provide a default architecture of a previously trained ResNet152, but it requires erasing its last fully connected layer and creating a new dense layer of a single neuron. Once trained, the neural network should output a continuous value in the interval [1,2] corresponding to the fractal dimension of the basin of attraction of the image input into the network.\\

The training of the network starts with random initial weights and biases. These parameters are tuned afterwards with the training set of basins of attraction of the Duffing oscillator, as explained in section~\ref{FractalDimension}. During this phase, the adaptive moment estimation method \textit{Adam}~\cite{kingma2014adam} optimizes the weights and biases with the hyperparameters: \textit{learning rate} $ = 0.001 $, $ \beta_1 = 0.9 $, $ \beta_2 = 0.999 $, $ \epsilon = 10 ^ {- 7} $, and $ amsgrad $ = False.\\

Finally, we use the function \texttt{imagedatagenerator} to feed images from the dataframe to the network. This function allows to present images to the neural network and to label them accordingly. Furthermore, it facilitates the recognition of useful patterns with the application of data augmentation techniques, such as rotation, shear, or zoom. In this work, the modifications made by this function to the training images have been limited to the random rotation of a set of randomly chosen images from the dataset, and scaling down the resolution of each input image into $ 224 \times 224 $ pixels to match the input resolution of the ResNet152 architecture. Although this last modification might introduce complications when working with highly fractalized basins (due to the fact that this type of basins have convoluted boundaries), this is a constrain imposed by the architecture.\\

With this setup, we train the neural network with images of basins of attraction from the Duffing oscillator to produce a CNN, from which we can predict the fractal dimension. The main achievement is that the computation of the fractal dimension of such images is faster and with a similar accuracy than by using the conventional method.

\section{Training set: basins and their fractal dimension}
\label{FractalDimension}

As for the image set, we have used $5511$ images corresponding to basins of the Duffing oscillator defined by
\begin{equation}
\ddot{x}+\delta \dot{x}+\alpha x+\beta x^{3}=\gamma \cos \omega t.
\end{equation}

\begin{figure}
\centering
\includegraphics[scale=0.45]{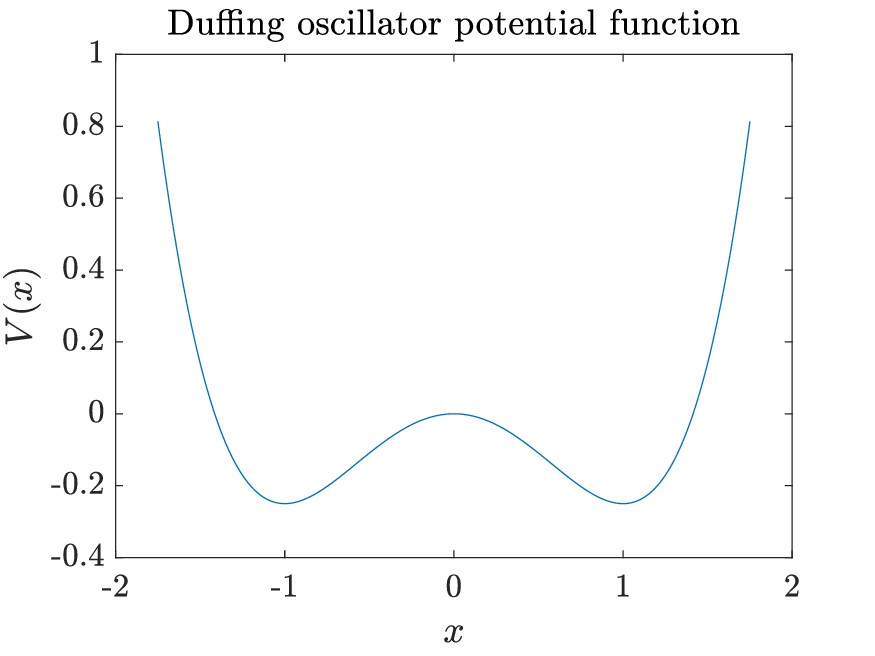}
\caption{Potential function for the Duffing oscillator, the function shows a double well where it is intuitive to think of at least three possible outcomes for a particle inside the potential. $(a)$ The particle oscillates in the left well without leaving it, $(b)$ the particle oscillates in the right well without leaving it, $(c)$ the particle oscillates between both wells not settling into anyone. When the motion of the particle in the system is chaotic the basin of attraction shows fractal basin boundaries. However, in case of non chaotic motions, the basins of attraction will show a smooth boundary.}
\label{doublewell}
\end{figure}

For this work, we set the parameters of this nonlinear oscillator to $ \delta = 0.15 $, $ \alpha = -1 $, and $ \beta = 1 $, while $ \gamma$ has been modified between the values $[0.1,0.5]$, and $\omega$ between $[0.2,2.5]$. This choice of parameters in the differential equation describes the motion of a unit mass particle in a double well potential of the form $V(x) = \frac{x^4}{4} - \frac{x^2}{2}$ as shown in Fig.~\ref{doublewell}, where the particle also experiences a linear drag with the environment $(\delta)$ and a external oscillatory forcing ($\gamma \cos \omega t$). \\

The basins of attraction have been computed numerically using a standard RK45 integrator based on the algorithm used in \textit{Dynamics: Numerical Explorations}~\cite{nusse1998dynamics}. We compute such basins setting the external forcing value and frequency to a parameter in the range mentioned above, and testing for $10^6$ initial conditions of $x$ and $\dot{x}$, coloring each pixel corresponding to the asymptotically behavior of the initial condition and therefore obtaining an image of $1000 \times 1000$ pixels.\\

\begin{figure}
\centering
	\begin{subfigure}[]
		\centering
		\includegraphics[scale=0.22]{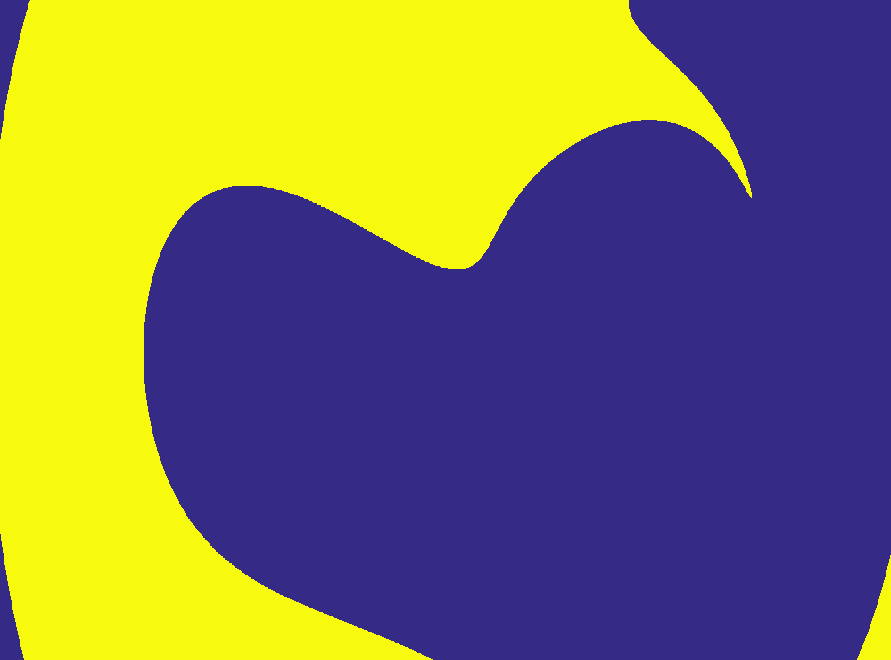}
		\label{1.04}
	\end{subfigure}
	\begin{subfigure}[]
		\centering
		\includegraphics[scale=0.22]{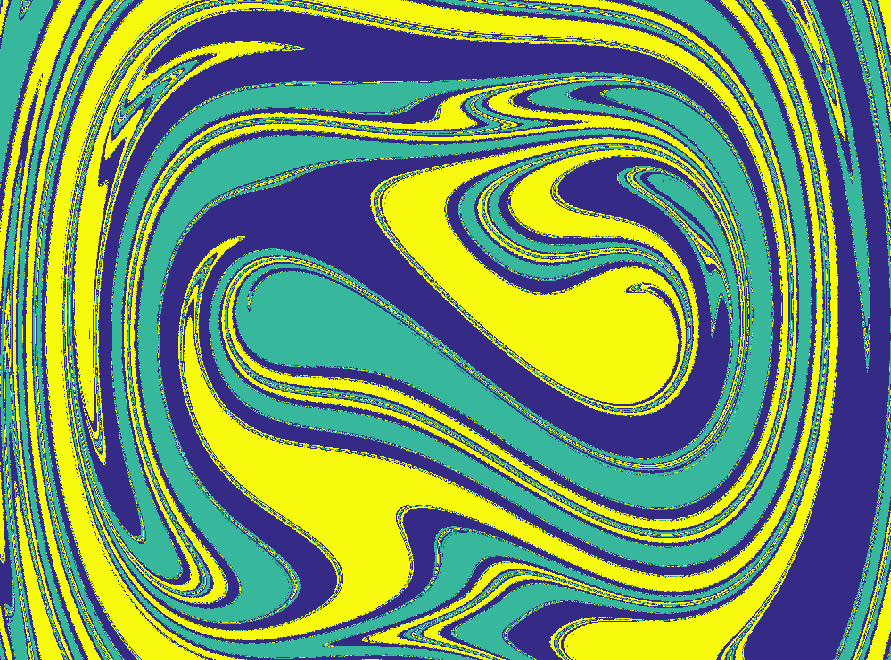}
		\label{1.85}
	\end{subfigure}
	\begin{subfigure}[]
		\centering
		\includegraphics[scale=0.22]{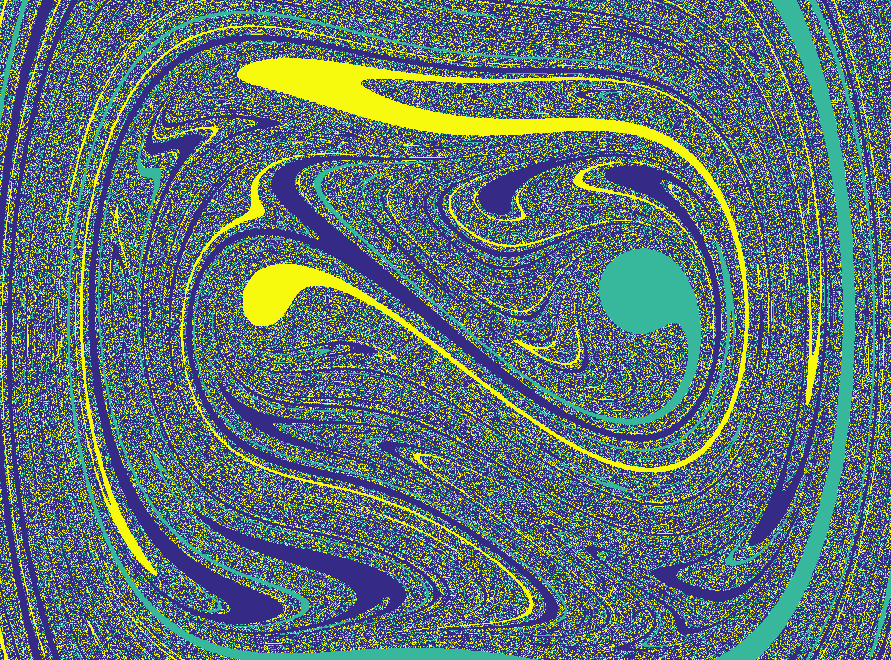}
		\label{1.98}
	\end{subfigure}
\caption{Examples of basins of attraction from the Duffing oscillator used in the training of the CNN. Basin (a) has a fractal dimension $F_{dim} = 1.04$, while (b) has $F_{dim} = 1.85$, and finally (c) has $F_{dim} = 1.98$. Intuitively, it is easy to see that the less fractalized basin is (a) since its basin boundary is quite smooth, while (b) seems more complex and therefore has a larger fractal dimension, and finally (c) is even more complex and hence has a even higher fractal dimension. Therefore our goal is to train the CNN to classify correctly these basins.}
\label{BasinImages}
\end{figure}

For this parameter set we obtain basins of attraction with a fractal dimension which varies between 1 and 2 depending on the fixed $\gamma$ and $\omega$ values. Figure~\ref{BasinImages} shows examples of basins with which the neural network trains. These images show that basins of attraction with smooth boundaries have a smaller fractal dimension than those who have convoluted boundaries. Thus, our goal is to train the CNN to recognize these patterns and estimate the fractal dimension of the basins.\\

After computing the basin set, each basin is labeled with the fractal dimension associated with the basin boundaries. To carry out the labeling, we have used the box-counting method proposed by McDonald et al.~\cite{mcdonald1985fractal}. Our implementation of the algorithm computes the final state unpredictability of all the boundaries present in the computed basins of attraction. The neural network will be trained on the average values of the fractal dimension and will therefore be able to predict as accurately as those average values that we will consider as our ground truth.\\

To minimize the estimation error of the box counting dimension, and therefore train our CNN to be as accurate as our ground truth, the result of the fractal dimension for a basin of attraction has been averaged over $10$ trials. The standard deviation of this sample is taken as the estimation error. For each trial, we have drawn at random $350000$ square boxes of sizes ranging from $10$ to $100$ pixels in steps of $10$ pixels in order to cover the entire basin.\\

The total $5511$ images are separated into three randomly distributed sets. The training set for the neural network amounts to $70 \%$ of these images. $10 \%$ of the total is kept for improving the results through the validation step. And finally, the test set checks the prediction accuracy of the network with the remaining $20 \%$ of the images.\\

\begin{figure}[]
\centering
\includegraphics[scale=0.5]{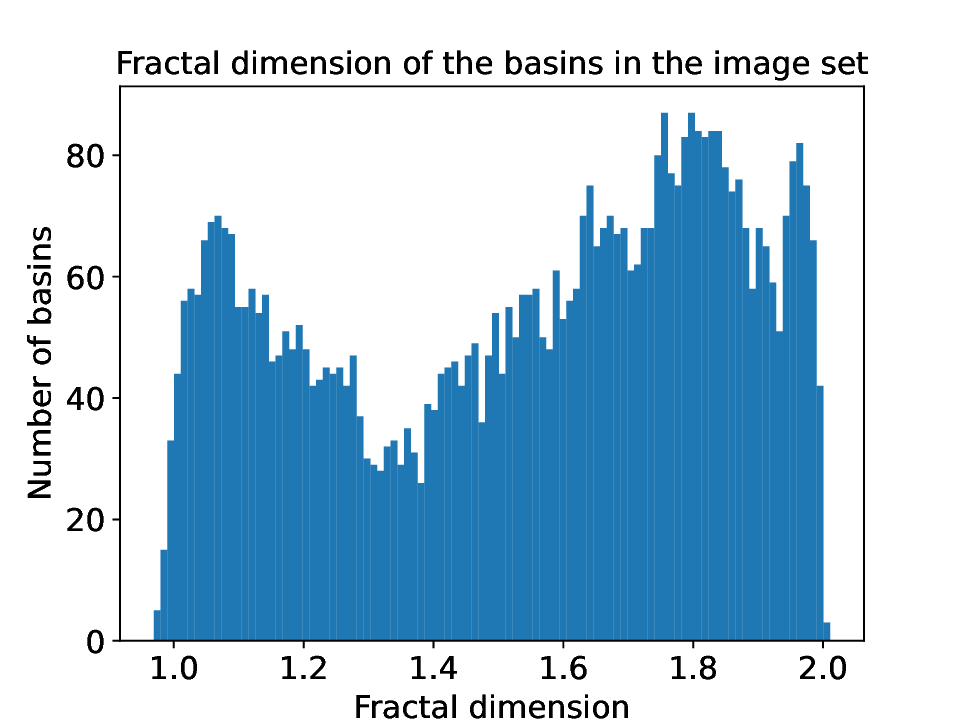}
\caption{Histogram that shows the frequency of the fractal dimension associated to every basin of the image set. The histogram is divided in $100$ bins and it is obvious that the neural network is biased to train with higher dimensional basins.}
\label{GTBasins}
\end{figure}

The choice of the parameter set for the basins of the Duffing oscillator has consequences on the distribution of the fractal dimension. The values are not uniformly distributed as the histogram of Fig.~\ref{GTBasins} shows. This uneven input distribution produces biases in the training of the neural network ~\cite{Kim2019LearningNT} since the basins of attraction with higher fractal dimensions are more frequent in the training set.\\

The loss function that minimizes the error of prediction of the ResNet152 has been established as \textit{mse} (mean square error). In addition, the last layer of the architecture is a fully connected layer projected to a single output node with a linear activation function. It is a suitable configuration for regression problems.\\

\begin{figure}[htb]
\centering
\includegraphics[scale=0.52]{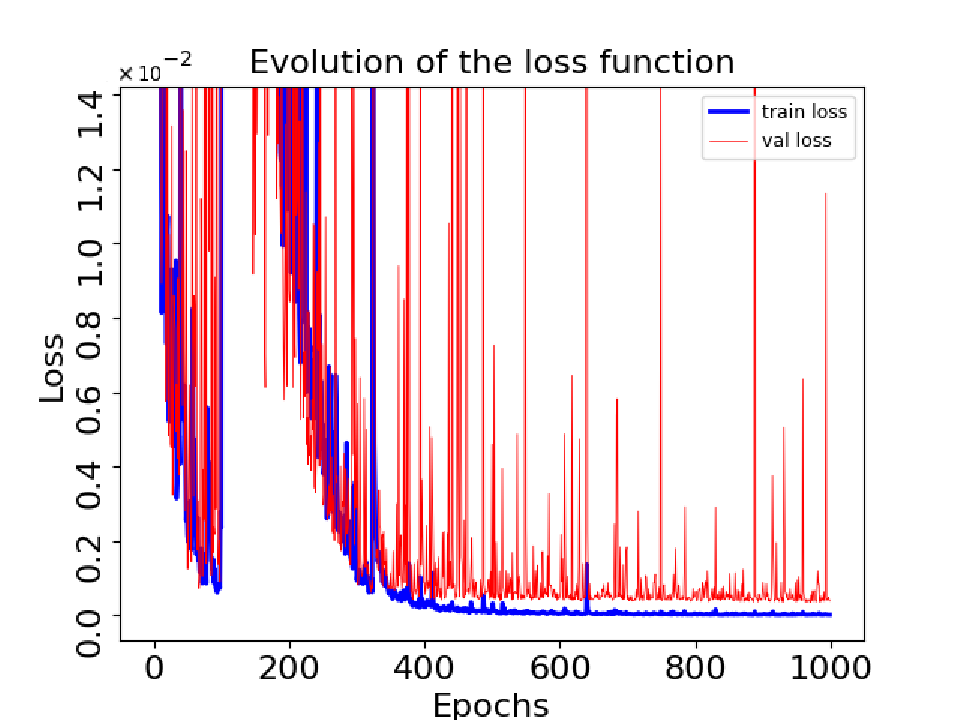}
\caption{Evolution of the loss function \textit{mse} during the training of the neural network for estimating the fractal dimension of basins of attraction. Images of basins from the Duffing oscillator have been used to train the network, using 3857 as the training set, and 552 as the validation set. It can be observed how the loss functions for both sets tend to 0, so we conclude that the neural network is capable of subtracting the features that contribute to the estimation of the fractal dimension in the training and validation set.}
\label{regression.loss}
\end{figure}

The training of the neural network for the estimation of the fractal dimension has been done in $1000$ iterations (epochs). The evolution of the loss function as a function of each iteration is shown in Fig.~\ref{regression.loss}, where it can be seen how the loss function for both the training set and the validation set tend to a value close to zero. So we conclude that the neural network learns to accurately predict both the training and validation set of images of basins of attraction.

\section{Predictions and performance}
\label{Results}

For the testing set, we used $1102$ images from basins of attraction of the Duffing oscillator on random forcing parameters from the range explained in Sec.~\ref{FractalDimension}. We can measure the accuracy of the prediction of the neural network on the testing set compared to its labeling (what we consider our ground truth). To achieve that we use the root mean square error formula
\begin{equation}
\hat \sigma=\sqrt{\frac{\sum_{t=1}^{N}\left(\hat{x}_{t}-x_{t}\right)^{2}}{N}},
\end{equation}
where $\hat{x_t}$ is the predicted value, $x_t$ the expected value, and $N$ the number of observations.\\

The obtained value of the root mean square error for the Duffing oscillator test set is $\hat \sigma = \pm 0.012 $. So the network was able to estimate the fractal dimension of those $1102$ basins with an error of this order on average, taking approximately $0.4$ seconds to estimate the fractal dimension of each image on the dedicated hardware specified on Sec.~\ref{ArchitectureChapter}. Then we have compared this estimation method to our labeling method. For the box-counting method proposed by McDonald et al.~\cite{mcdonald1985fractal}, the conventional method took $4.7$ seconds approximately per image to estimate the fractal dimension of each basin with an standard deviation of $\sigma = \pm 0.012$. This proves that our neural network is able to estimate fractal dimensions approximately $10$ times faster than the conventional method.\\

\begin{figure}[]
\centering
\includegraphics[scale=0.56]{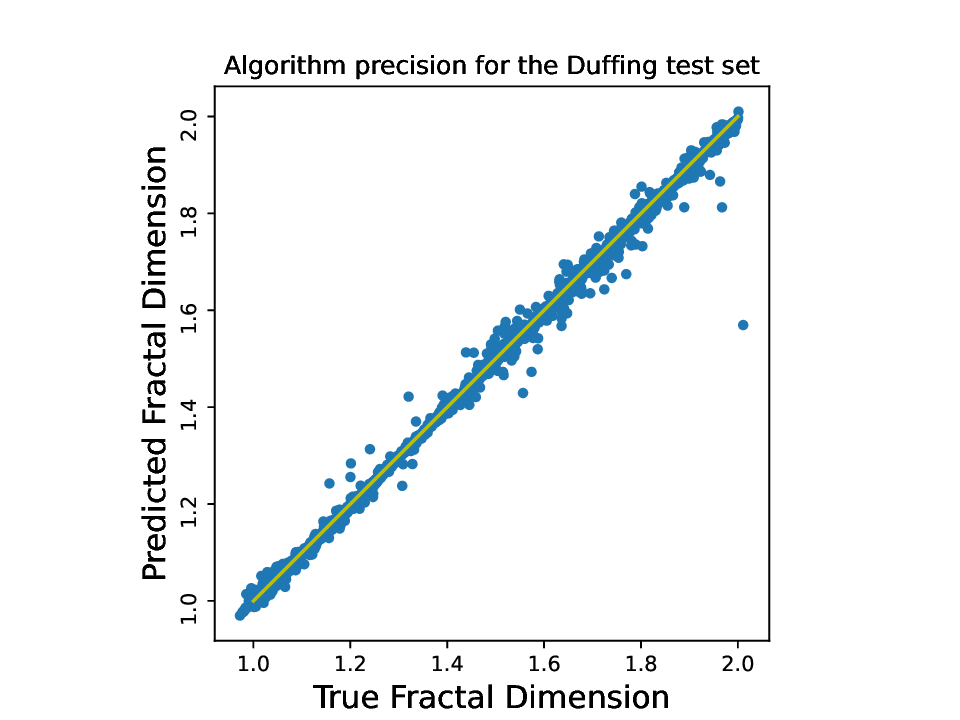}
\caption{Comparison for each image in the test set of the fractal dimension considered as ground truth (x-axis), and the fractal dimension that the algorithm predicts (y-axis). It can be seen how the algorithm is able to predict with a small margin of error the fractal dimension of basins of attraction from the Duffing oscillator, obtaining a linear fit with a Pearson correlation coefficient of $R^2 = 0.998$.}
\label{LinearRegression}
\end{figure}

\begin{figure}[htb]
\centering
\includegraphics[scale=0.5]{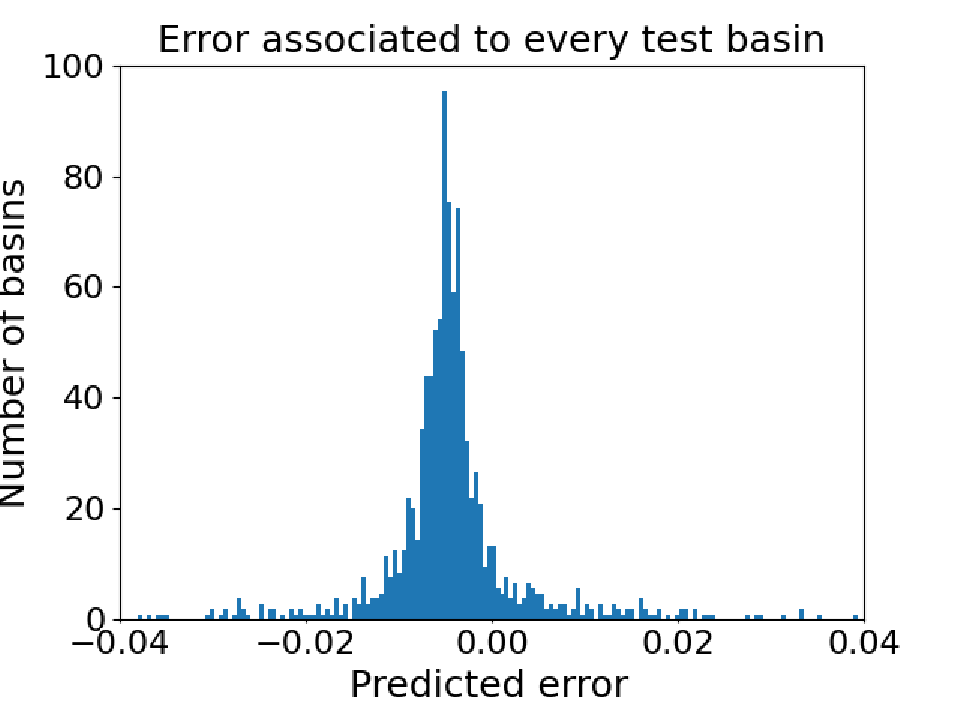}
\caption{This histogram shows the error associated with each of the predicted basins with respect to its labeling value. It can be observed how the majority of predicted basins have an error of less than \mbox{$\pm$ 0.01}, so we can conclude that the algorithm is quite accurate.}
\label{LinearHist}
\end{figure}

On the other hand, Fig.~\ref{LinearRegression} shows a linear regression of the comparison between the fractal dimension values of the labeling (our ground truth) and the fractal dimension that the network predicts for this same set. It can be seen how most of the basins are predicted accurately, although there is always some error between our ground truth and the prediction from the network. This error is shown in Fig. ~\ref{LinearHist} as a histogram with $100$ bins, where we classify basins depending on the deviation of the predicted fractal dimension and the fractal dimension used as a ground truth.\\

When considering the shape of the histogram in Fig. ~\ref{LinearHist}, it is clear that it approximates to a Gaussian distribution with a mean of approximately $-0.005$. This bias in the mean is due to two factors: first, the lack of uniformity in the distribution of basins with different fractal dimensions, since as shown in Fig.~\ref{GTBasins} the neural network trains mainly with basins of high fractal dimension producing a bias in the neural network. Secondly, the basin of attraction images were initially computed with a resolution of $1000\times1000$ pixels. When introduced to the network, the image is scaled down to $224 \times 224$ pixels to match the input of the ResNet152 architecture, therefore producing an error when estimating the fractal basins.

\section{Conclusions}
\label{Conclusions}
We have successfully programmed a neural network that estimates the fractal dimension of a basin of attraction of the Duffing oscillator. Our neural network is capable of predicting fractal dimensions for this system in the parameter space $\gamma$ = $[ 0.1,0.5]$, and $\omega$ = $[0.2,2.5]$, while fixing $\delta = 0.15 $, $ \alpha = -1 $, and $ \beta = 1 $. We have obtained an average error on the estimation of $\hat \sigma = \pm 0.012$ in an average time of $0.4$ seconds. Note that the previous error is generally larger for basins of higher fractal dimension. As a consequence, this method becomes $10$ times faster than the conventional box-counting method.\\

With regards to the labeling of the images, the average error of the fractal dimension using the box-counting dimension algorithm ~\cite{mcdonald1985fractal} is $0.002$, which is a much lower value than the prediction error of the network. However, we must say that this labeling error depends on the number of trials for the box-counting computation. As such, the labeling error can be arbitrarily small and as we have observed.\\

One way to improve these results would be to increase the size and homogeneity of the training and validation sets in the fractal dimension, since it would allow the network to minimize the error produced on the estimation of the lower dimensional basins. On the other hand, the estimation of the higher dimensional basins requires the neural network to visualize all the details in the image of a basin, so a network with an architecture that inputs an image of bigger scale would be able to accurately predict basins with a high fractal dimension.\\

Although the neural network does not achieve the same average error as the conventional method, we conclude that the trade-off between the computing speed and the loss of precision, is still in favor of the deep learning technique. Note that as shown earlier, if we set the conventional algorithm to have the same average error as the neural network predictions, we observe that the neural network predicts the fractal dimension of a basin approximately $10$ times faster than the conventional method.\\

\section*{ACKNOWLEDGMENTS}
This work has been financially supported by the Spanish State Research Agency (AEI) and the European Regional Development Fund (ERDF) under Project No.~PID2019-105554GB-I00.


%

\end{document}